\documentclass[draft]{agujournal2019}
\usepackage{apacite}
\usepackage{url} 
\usepackage[inline]{trackchanges} 
\usepackage{soul}
\usepackage{bm}
\usepackage{amsmath}
\usepackage{enumitem}
\draftfalse

\journalname{JGR: Planets}

\begin{document}

\title{Particle in cell simulation on mode conversion of Saturn's 20 kHz narrowband radio emission}
%
%
\authors{Zhoufan Mu\affil{1}, Yao Chen\affil{1,2,3}, Tangmu Li\affil{1}, Sulan Ni\affil{4}, Zilong Zhang\affil{2,3}, and Hao Ning\affil{1}}

\affiliation{1}{Institute of Frontier and Interdisciplinary Science, Shandong University, Qingdao, Shandong 266237, PR China}
\affiliation{2}{Shandong Key Laboratory of Space Environment and Exploration Technology, Shandong University, Shandong 264209, PR China}
\affiliation{3}{Institute of Space Sciences, Shandong University, Shandong 264209, PR China}
\affiliation{4}{School of Physics and Electronic Information, Yantai University, Yantai 264005, PR China}

\correspondingauthor{Yao Chen}{yaochen@sdu.edu.cn}



\begin{keypoints}
\item We conduct fully-kinetic 2D3V particle-in-cell simulation of the mode conversion process underlying the Saturn's 20 kHz narrowband radiation.
\item We use the wave-pumping method to initiate the Z-to-O mode conversion induced by the density gradient.
\item We identify the Z mode-converted electromagnetic radiation and estimate the mode conversion rate to be ~10-20\%.
\end{keypoints}

%
%

%
%


\begin{abstract}

The Z-to-O mode conversion at the density gradient is the prevailing mechanism of narrowband (NB) radio emission in planetary magnetosphere. Most previous numerical models were for NB emission observed in the Earth magnetosphere, using the cold plasma fluid approximation that excluded any kinetic effect. Here we investigate the Z-to-O conversion process underlying the Saturn’s 20 kHz NB emission, using the fully-kinetic and electromagnetic particle-in-cell (PIC) simulation. We simulate the whole process starting from the pumping of the Z mode, to its propagation and reflection, and further conversion into the O mode radiation. The energy conversion rate of the Z-to-O process is estimated to be ~10-20\%. This provides the first quantitative estimate of such rate with PIC simulations.
\end{abstract}

\section*{Plain Language Summary}

Saturn emits 20~kHz narrowband radio signals in regions where plasma density changes abruptly. Using advanced computer simulations, we discovered how trapped waves can convert to radio waves that can escape into space: When these ``trapped waves" encounter areas of increasing density—much like light propagating into lens—their path curves till they align with the magnetic field. This alignment allows them to transform into radio signals that can be detected on Earth. Our key finding is that 10-20\% of the trapped wave energy converts into these escaping signals. This explains Saturn's narrowband radio emissions and helps scientists decode similar signals observed near other planets by missions like Cassini and Juno, improving our understanding of planetary space environments.

%
%

%


%
%
%
%

\section{Introduction} \label{sec:intro}

Plasma emission refers to coherent radiation from plasmas at frequencies close to the fundamental plasma frequency ($\omega_\mathrm{pe}$) and/or its harmonics \cite{Ginzburg_1958}. Two processes have been proposed to account for such emissions. The first one is the standard beam-driven plasma emission that is a nonlinear and multi-step process starting from the bump-on-tail instability driven by beam of energetic electrons. The instability excites Langmuir waves which interact with density fluctuations or the ion acoustic wave non-linearly to generate the escaping O mode wave (i.e., the fundamental branch). These Langmuir waves also interact with the induced backward-propagating Langmuir waves to generate the harmonic radiation at 2~$\omega_\mathrm{pe}$ \cite{Ginzburg_1958,Melrose_1980,Cairns_1985,Robinson_1994,Robinson_1994a}

The second plasma emission process is given by the mode conversion from Z to O at density gradient. Such process has been generally used to explain the narrowband (NB) radiation observed from planets, or the nonthermal continuum (NTC) radiation observed in the Earth’s magnetosphere. \citeA{Gurnett_1975} were the first to report the latter radiation using IMP-6 data. Subsequent observations revealed similar emissions, which have been termed as the NB radiation, from the magnetospheres of Saturn, Jupiter and its moon Ganymede \cite{Kurth_1981,Kurth_1997,Kurth_2022,Gurnett_1975,Gurnett_1981,Gurnett_1983,Gurnett_1996,Grimald_2008,Grimald_2010}. Such radiation shares the following common characteristics \cite{Menietti_2012,Decreau_2015}: (1) having multiple harmonic structures; (2) being closely associated with intense Upper Hybrid (UH) or Z mode (i.e., the slow extraordinary mode); (3) originating from regions with steep density gradients such as plasma sheet boundaries or planetary rings.

\citeA{Jones_1980,Jones_1987} proposed the Z-to-O mode conversion process in density gradient to explain the plasma emissions from planets. This process has three stages: (1) High-energy electrons excite multiple UH oscillations with large wavenumbers through the double plasma resonance process along the structure with density gradient; (2) As these UH oscillations propagate into denser regions, their wavenumber decreases till they evolve into the electromagnetic Z mode; (3) The Z mode gets reflected at a certain location. The reflected Z mode propagates into the so-called Ellis window when its frequency $\omega$ is close to the local plasma oscillation frequency $\omega_\mathrm{pe}$ \cite{Ellis_1956,Ellis_1962}. Within the window both Z and O modes propagate parallel to the background magnetic field with identical frequency and wavenumber. This leads to the mode conversion process. Recent studies also employed the Z-to-O mode conversion to explain solar radio bursts \cite{Sakai_2005,Willes_2001,Krafft_2024b,Krafft_2024c,Krafft_2024d}.

\begin{figure}
\noindent
\includegraphics[width=0.8\textwidth]{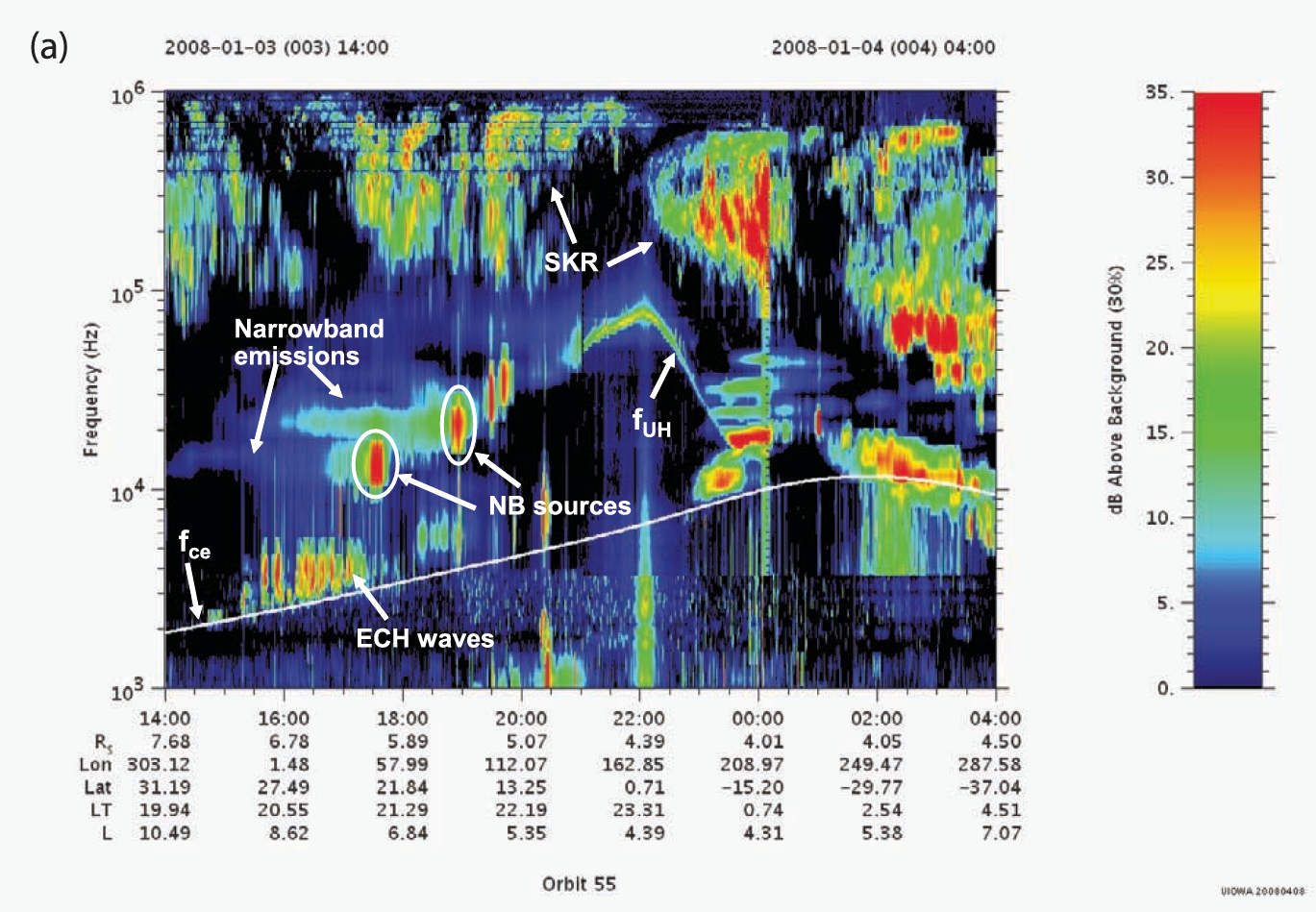}
\caption{Spectrogram of the Cassini-RPWS observations on 04/01/2008 during its crossing of the source region of the 20~kHz NB. The detection of Saturn Kilometric Radiation (SKR), Electron Cyclotron Harmonics (ECH), narrowband radiation (NB) and their sources are annotated in white. The white line represents the electron cyclotron frequency $f_{ce}$. The x-axis corresponds to the Cassini observations time period, radial distance (Rs), longitude (Lon), latitude (Lat), local time (LT) and L-shell value (L). The y-axis corresponds to the Cassini-RPWS frequency channels from 1~kHz to 1~MHz, distributed on a logarithmic scale. This Figure is adapted from Figure 1 of \citeA{Ye_2009}).}
\label{fig:Cassini}
\end{figure}

Figure~\ref{fig:Cassini} from \citeA{Ye_2009} displays the 20 kHz NB and the associated electrostatic disturbances detected by Cassini-RPWS while crossing the boundary layer of Saturn's plasma sheath on 04/01/2008. Electrostatic UH signals showed a sharp increase in frequency, indicating a density gradient at the sheath boundary. According to \citeA{Ye_2009}, the radiation has multiple bands with left-handed polarization, within 3 to 70~kHz. They also demonstrated that the weak NB signals are linked to the strong electrostatic source, indicating that the NB radiation originates from the intense UH/Z mode waves. According to the intensity ratio indicated by the color scale in Figure~\ref{fig:Cassini}, the energy conversion rate from UH/Z mode to NB radiation is ~1\%. These observations support the NB origin of the UH/Z-to-O conversion process due to the density gradient \cite{Ye_2009,Ye_2010,Menietti_2009}.

\citeA{Kim_2007,Kim_2008} simulated mode conversion in warm plasma under typical solar wind conditions, using an electron-fluid model with density gradient. Their study focused on the conversion of Langmuir waves into electromagnetic (X/O mode) radiation, and estimated the maximum efficiency of mode conversion from the incident Langmuir wave to the electromagnetic wave. \citeA{Kalaee_2009} injected UH waves obliquely to the magnetic field with a 2D fluid simulation, setting the width of the density gradient to be ~5 times the wavelength of the UH mode. They showed the occurrence of the Z-to-O mode conversion, and estimated the Poynting flux ratio of O mode to initial UH mode to be ~50\%. In a series of studies, the authors investigated how different propagation angles and density gradient widths affect the conversion efficiency \cite{Kalaee_2014,Kalaee_2020}.

PIC simulations self-consistently solve the equations of particle motion and those of the electromagnetic fields, thus the method has been used to verify the nonlinear plasma emission process \cite{Thurgood_2015,Henri_2019,Chen_2022,Zhang_2022}. For instance, \citeA{Zhang_2022} conducted PIC simulations within a uniform domain with a large number of macroparticles, and estimated the conversion rate of energy from the beam of energetic electrons to be $\sim5*10^{-6}$ for the fundamental emission and $\sim4*10^{-5}$ for the harmonic emission. Studies by \citeA{Krafft_2021} have investigated the effect of density fluctuations on the plasma emission process. They concluded that the ratio of harmonic emission energy to Langmuir wave turbulence energy exceeds $\sim10^{-3}$ in inhomogeneous plasma and $\sim10^{-4}$ in homogeneous plasma, and found that density fluctuations tend to enhance the growth rates of harmonic emission.

\citeA{Horky_2018,Horky_2019} conducted PIC simulations of the NTC radiation under conditions of the Earth's magnetosphere by injecting ring-distributed energetic electrons into regions of increasing density. They suggested the obtained O mode radiation originate from the conversion of the ring-excited Bernstein mode due to the density gradient, though no obvious signature of electromagnetic waves was observed in the post-conversion dispersion diagram (see Figure 7 of \citeA{Horky_2018}). This implicates the necessity of further study on the mode conversion process induced by density gradient.

In this article, we use fully kinetic and electromagnetic PIC simulations to study the Z-to-O mode conversion process induced by density gradient. We also estimate the energy conversion rate of the process. In Section \ref{sec:meth&para} we present the numerical method and parameter setup. The simulation result is presented in Section~\ref{sec:results}, and Section ~\ref{sec:conclusion} and ~\ref{sec:discussion} present our conclusion and discussion, respectively.

\section{Numerical method and parameter setup} \label{sec:meth&para}

We used the Vector-PIC (vPIC) open-source code released by the Los Alamos National Laboratory \cite{Bowers_2008,Bowers_2009}. The code applies a second-order, explicit leapfrog algorithm to resolve the particle motion and a second-order finite-difference time-domain solver to solve the full Maxwellian equations of the electric-magnetic fields. The simulations are 2D3V, i.e., with two spatial dimensions in the XOZ plane and three velocity components, with periodic boundary conditions.

We set the thermal velocity ($v_{the}$) of both electrons and protons to be 0.01c \cite{Menietti_2019}. We set the background magnetic field along the z-axis (i.e., the parallel direction, $\vec B_0 = B_0 \hat{e}_z$), and the wave vector $\vec{k} = k_x \hat{e}_x + k_z \hat{e}_z$, where $k_{\parallel} = k_z, k_{\perp} = k_x$. The simulation domain spanned $L_x = L_z = 18~c/\Omega_\mathrm{ce}$, with a grid size of $2048 \times 2048$. We set the time step to be $\Delta t = 0.0217~\omega_\mathrm{pe}^{-1}$ and ran the simulation for $4000~\omega_\mathrm{pe}^{-1}$. We used the realistic proton-to-electron mass ratio (mp/me = 1836), and employed 1000 particles of each species in each grid. The total number of particles is $8.4 \times 10^{9}$.

According to Figure~\ref{fig:Cassini}, the ratio ${\omega_\mathrm{pe}}/{\Omega_\mathrm{ce}}$ is 5. We determine the O mode radiation frequency from the figure to be $\omega = 5.03~\Omega_\mathrm{ce}$. Following \citeA{Kalaee_2009}, we solved the cold-plasma magnetoionic dispersion relation together with the Snell’s law (see Figure~\ref{fig:Z mode solution}). The density gradient is parallel to magnetic field ($\vec{B}$), so both $\omega_\mathrm{pe}$ and the refractive index $N$ vary perpendicular to $\vec{B}$. According to the Snell’s law, both $\theta$ (the angle between $\vec{k}$ and $\vec{B}$) and $k_x$ vary with $N$, while $k_z$ remains constant. The Z mode propagates into the Ellis window with $\omega_\mathrm{pe}\sim\omega$, where the mode conversion occurs.

\begin{figure}
\noindent
\includegraphics[width=0.8\textwidth]{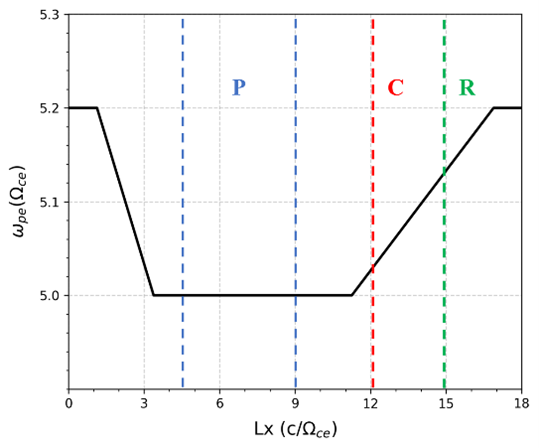}
\caption{Slice of background plasma density along the $L_x$ direction (normalized by $c/\omega_\mathrm{pe}$), expressed as plasma frequency $\omega_\mathrm{pe}/\Omega_\mathrm{ce}$. The density gradient increase from 5.0 to 5.2~$\Omega_\mathrm{ce}$. Two blue lines demarcate the Z mode pumping zone ``P" ($L_x=4.5-9~c/\Omega_\mathrm{ce}$). Green line marks the theoretical return zone ``R" ($\omega_\mathrm{pe}=5.13~\Omega_\mathrm{ce}$) and red line marks the Z-to-O mode conversion zone ``C" ($\omega_\mathrm{pe}=5.03~\Omega_\mathrm{ce}$).}
\label{fig:density model}
\end{figure}

Figure~\ref{fig:density model} presents the setup of the density gradient that is perpendicular to $B_0\cdot e_z$. We set $\omega_\mathrm{pe}$ increases from 5~$\Omega_\mathrm{ce}$ to 5.2~$\Omega_\mathrm{ce}$ over a width of ~5.6~$c/\Omega_\mathrm{ce}$. With the method presented in~\ref{sec:LMCT}, we deduced the initial wavenumber of the Z mode to be $k_0 = 6.0~\Omega_\mathrm{ce}/c$, and its propagation angle $\theta_\mathrm{0} = 70.6^{\circ}$. With the method presented in~\ref{sec:pumping}, we pumped the Z mode into the uniform region (blue lines in Figure~\ref{fig:density model}). According to the linear theory of cold-plasma, the Z mode will get reflected at the return point (green line) where $\omega_\mathrm{r} = 5.13~\Omega_\mathrm{ce}$, it partially converts to the O mode when it reaches the Ellis window (red line) where $\omega_\mathrm{c} = 5.03~\Omega_\mathrm{ce}$.

\section{PIC simulations of mode conversion} \label{sec:results}

This section presents the complete process of mode conversion, starting from the pumping of Z mode to its further reflection and mode-conversion. According to Figure~\ref{fig:fields}, we pumped the Z mode in the uniform region with $k = 6.0~\Omega_\mathrm{ce}/c$, $\omega = 5.03~\Omega_\mathrm{ce}$, and $\theta_0 = 70.6^{\circ}$. We split the whole process into four stages:

\begin{enumerate}[label=\arabic*)]
	\item From 0 to 400~$\omega_\mathrm{pe}^{-1}$ and then to 1300~$\omega_\mathrm{pe}^{-1}$, the Z mode propagates from the uniform to the nonuniform region, within the latter region $\theta_0$ decreases from $70^{\circ}$ to $\sim 30^{\circ}$.
	\item From 1300 to 1800~$\omega_\mathrm{pe}^{-1}$, the upper part of the Z mode is reflected at the return zone, where $\theta \sim 30^{\circ}$. Then it moves into the less-denser region with decreasing $\theta$. When $\theta\sim0$ or the mode propagates almost parallel to $\vec B_0$, it reaches the conversion zone while propagates into the uniform region.
	\item From 1800 to 2000~$\omega_\mathrm{pe}^{-1}$, with the continuous entry of the Z mode into the return zone, an interference signal appears with the incident and reflected Z modes. According to Figure~\ref{fig:fields}, as more and more Z mode gets reflected, we observed continuous enhancement of parallel and quasi-parallel propagating waves, as a result of continuous mode conversion.
	\item From 2000 to 2700~$\omega_\mathrm{pe}^{-1}$, all the Z mode is reflected around the return zone. After that, the signal of the reflected Z mode dominates over that of the O mode (not shown here). In the following text, we analyzed the changes in wave number and energy during the wave mode conversion process, and calculated the conversion efficiency.
\end{enumerate}

\subsection{The mode conversion process}\label{sec:wavenumber}

A small-amplitude plane wave obeys the Snell’s law that means $k_\perp$ of the pumped Z mode decrease while $k_\parallel$ remains constant. The linear theory (see~\ref{sec:LMCT}) predicts the reflection take place at $\omega_\mathrm{r} = 5.13~\Omega_\mathrm{ce}$ where $\theta$ jumps from $\sim 30^{\circ}$ to $\sim -30^{\circ}$. The reflection occurs with a spatially-extended return zone with a length of $\sim 1~c/\Omega_\mathrm{ce}$.

Figure~\ref{fig:FFT} illustrates the wave intensity maps of electromagnetic (EM) fields in the wave–vector space, highlighting the wavenumber change of Z mode. Panel (a) shows the initial Z mode propagation in the uniform region from 0 to 400~$\omega_\mathrm{pe}^{-1}$, both $k$ and $\theta$ remain fixed to the prescribed values. In Panels (b) and (c), $E_x$ weakens with decreasing $\theta$, while $E_z$ dominates for $k_x\sim0$. Panel (d) for the post-conversion stage manifests the strong signal of the reflected Z mode, along with a weaker signal of parallel-propagating O mode. Figure~\ref{fig:dispersion relation} presents the $\omega-k$ dispersion relations for different propagation angles. Panel (e) shows a parallel-propagating O mode emerging during 1600–2400~$\omega_\mathrm{pe}^{-1}$, with the $E_x/E_z$ ratio consistent with the wave signatures observed in Figure~\ref{fig:FFT}.~\ref{sec:con eff} details the mode identification method used to deduce the occurrence of the Z-to-O mode conversion.

\subsection{Energy evolution and rate of mode conversion}\label{sec:energy}

Figure~\ref{fig:fields energy} displays the energy profiles of the six components of the EM field, normalized to their corresponding initial values, within the full domain (left) and the uniform region (right). During the propagation of the Z mode toward the density gradient (0 to 1700~$\omega_\mathrm{pe}^{-1}$), $E_x$ and total energy decrease throughout the domain. During the phase of mode conversion (1700 to 2100~$\omega_\mathrm{pe}^{-1}$), we observed $E_x$ weakening, $E_z$ strengthening, and $E_y$ remaining fixed in energy. Again, this is in line with the occurrence of mode conversion. From 2100 to 2900~$\omega_\mathrm{pe}^{-1}$, the Z mode reflects and gradually dissipates. In the uniform region, the Z mode is dominated by the $E_x$ component before 1700~$\omega_\mathrm{pe}^{-1}$. From 1700 to 2100~$\omega_\mathrm{pe}^{-1}$, $E_z$ increases significantly from $-1.8\times 10^{-5}$ to $-1.2\times 10^{-5}$, again, agreeing with the Z-to-O mode conversion. Figure~\ref{fig:fields energy} (left) shows the decrease of the total energy W of the pumped Z mode as it propagates. After 1300~$\omega_\mathrm{pe}^{-1}$, when the Z mode reaches the return zone, W remains nearly constant, while $E_x$ gets weaker, $E_y$ and $E_z$ get stronger in energy. This indicates a reduction of $\theta$ of the Z mode, as shown in Figure~\ref{fig:fields}.

We evaluated the wave energy within the black-dashed region marked in Figure~\ref{fig:fields}. Figure~\ref{fig:fields energy} reveals a decrease in Z mode energy during 0-1000~$\omega_\mathrm{pe}^{-1}$, indicating energy transfer from electromagnetic fields to plasma particles through wave-particle interactions. This energy decline ceased after 1000~$\omega_\mathrm{pe}^{-1}$. Consequently, we measured the pre-conversion Z mode energy ($W_\mathrm{E}^\mathrm{Z}$) at 1300~$\omega_\mathrm{pe}^{-1}$, while the post-conversion O mode energy ($W_\mathrm{E}^\mathrm{O}$) at 2000~$\omega_\mathrm{pe}^{-1}$. The conversion efficiency $\eta$ was calculated as the ratio of $W_\mathrm{E}^\mathrm{O}$ to $W_\mathrm{E}^\mathrm{Z}$, $\eta = \left( W_\mathrm{E}^\mathrm{O} \right) / \left( W_\mathrm{E}^\mathrm{Z} \right) = 17.5\%$.

\section{Discussion}\label{sec:discussion}

\begin{figure}
\noindent
\includegraphics[width=0.8\textwidth]{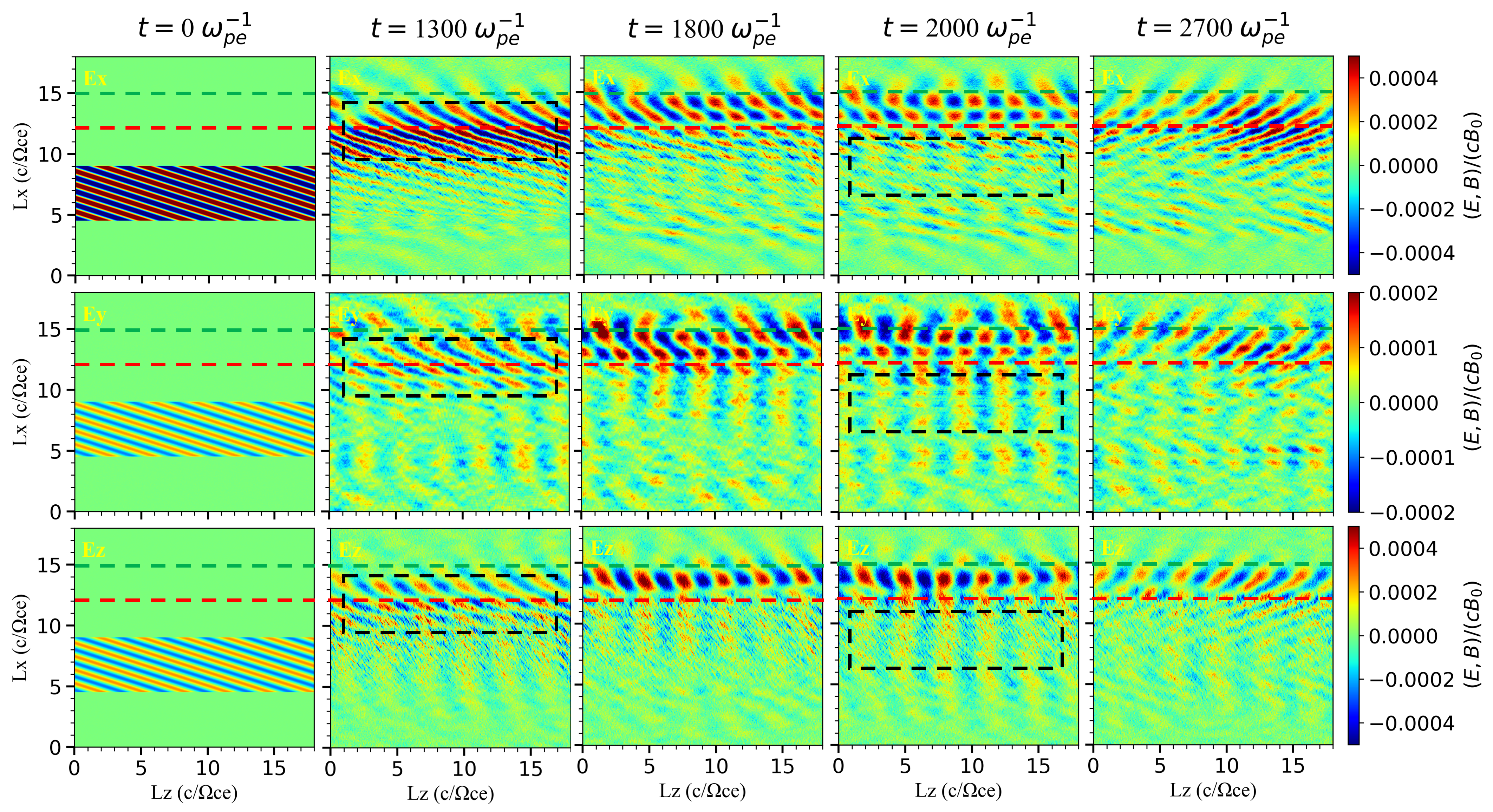}
\caption{Snapshot of electric field components ($E_x, E_y, E_z$) during mode conversion across five sequential phases: (1) initial state of pumped Z mode ($t=0~\omega_\mathrm{pe}^{-1}$); (2) propagation of Z mode in density gradient ($t=1300~\omega_\mathrm{pe}^{-1}$); (3) reflection of Z mode in return zone ($t=1800~\omega_\mathrm{pe}^{-1}$); (4) emergence of O mode ($t=2000~\omega_\mathrm{pe}^{-1}$); (5) post-conversion stage ($t=2700~\omega_\mathrm{pe}^{-1}$). Red and green lines mark conversion and return zones, black dashed box indicates the energy-calculation region (section~\ref{sec:energy}).}
\label{fig:fields}
\end{figure}

\begin{figure}
\noindent
\includegraphics[width=0.8\textwidth]{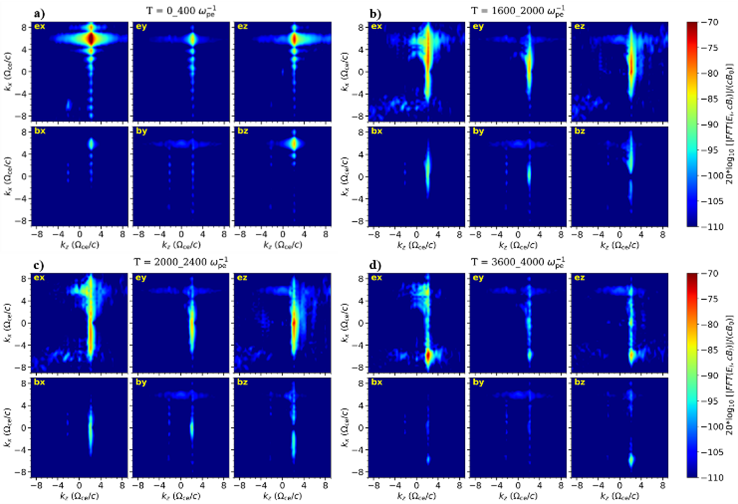}
\caption{Wave intensity maps of EM fields ($E_x, E_y, E_z, B_x, B_y, B_z$) in the wave–vector space ($k_x-k_z$), with wave number units of $\Omega_\mathrm{ce}/c$. Panel (a) corresponds to the initial Z mode in the uniform region ($0-400~\omega_\mathrm{pe}^{-1}$); Panels (b) and (c) correspond to the Z mode propagation in the non-uniform region (respectively $1600-2000~\omega_\mathrm{pe}^{-1}$ and $2000-2400~\omega_\mathrm{pe}^{-1}$); Panel (d) corresponds to the post-conversion stage with the reflected Z mode and the quasi-parallel O mode near $k_x\sim0$ (after $3600~\omega_\mathrm{pe}^{-1}$).}
\label{fig:FFT}
\end{figure}

\begin{figure}
\noindent
\includegraphics[width=0.8\textwidth]{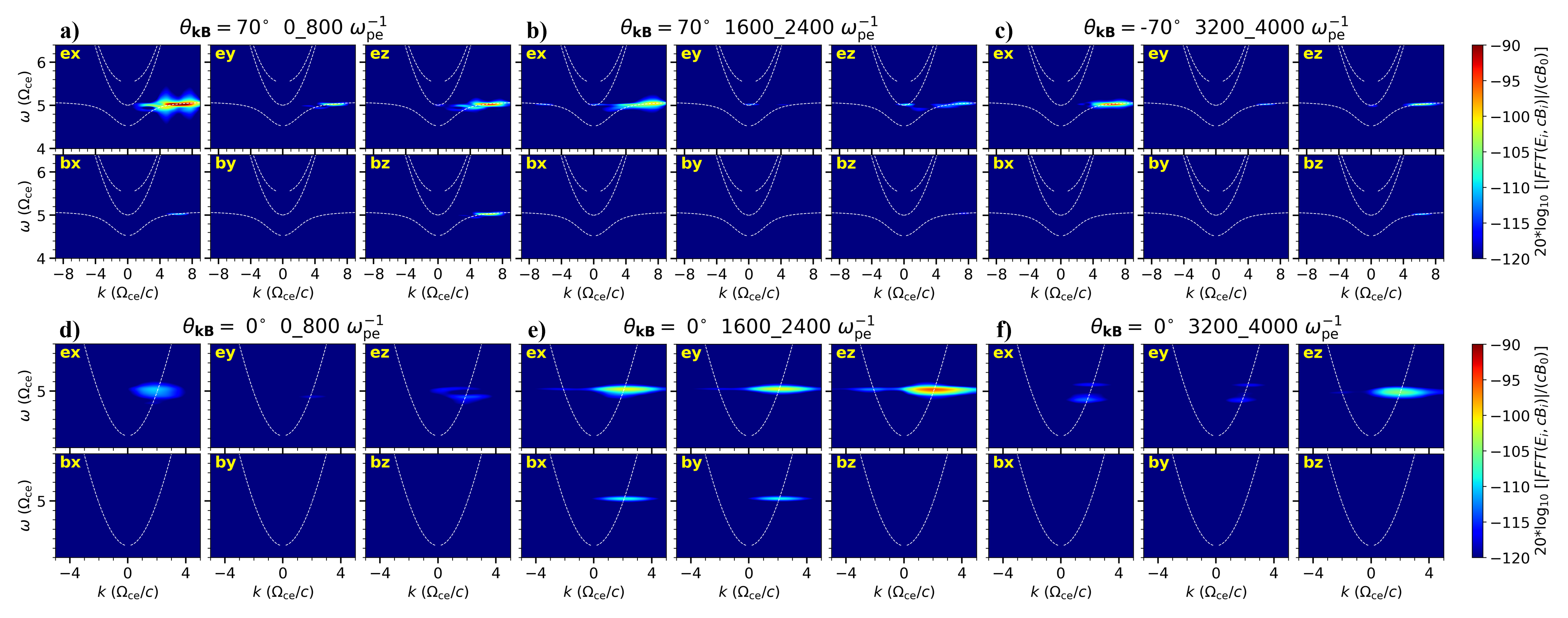}
\caption{Dispersion relations in uniform region for $\theta=70^{\circ} |\vec{k_{max}}|=9~\Omega_\mathrm{ce}/c$ (top) and $\theta=0^{\circ} |\vec{k_{max}}|=5~\Omega_\mathrm{ce}/c$ (bottom). Top row shows three temporal phases: (a) initial stage ($0-800~\omega_\mathrm{pe}^{-1}$); (b) mode conversion ($1600-2400~\omega_\mathrm{pe}^{-1}$); (c) post-conversion ($3200-4000~\omega_\mathrm{pe}^{-1}$). Bottom row shows quasi-parallel propagation of mode waves at corresponding times. White curves represent the theoretical dispersion in the uniform region.}
\label{fig:dispersion relation}
\end{figure}

\begin{figure}
\noindent
\includegraphics[width=0.8\textwidth]{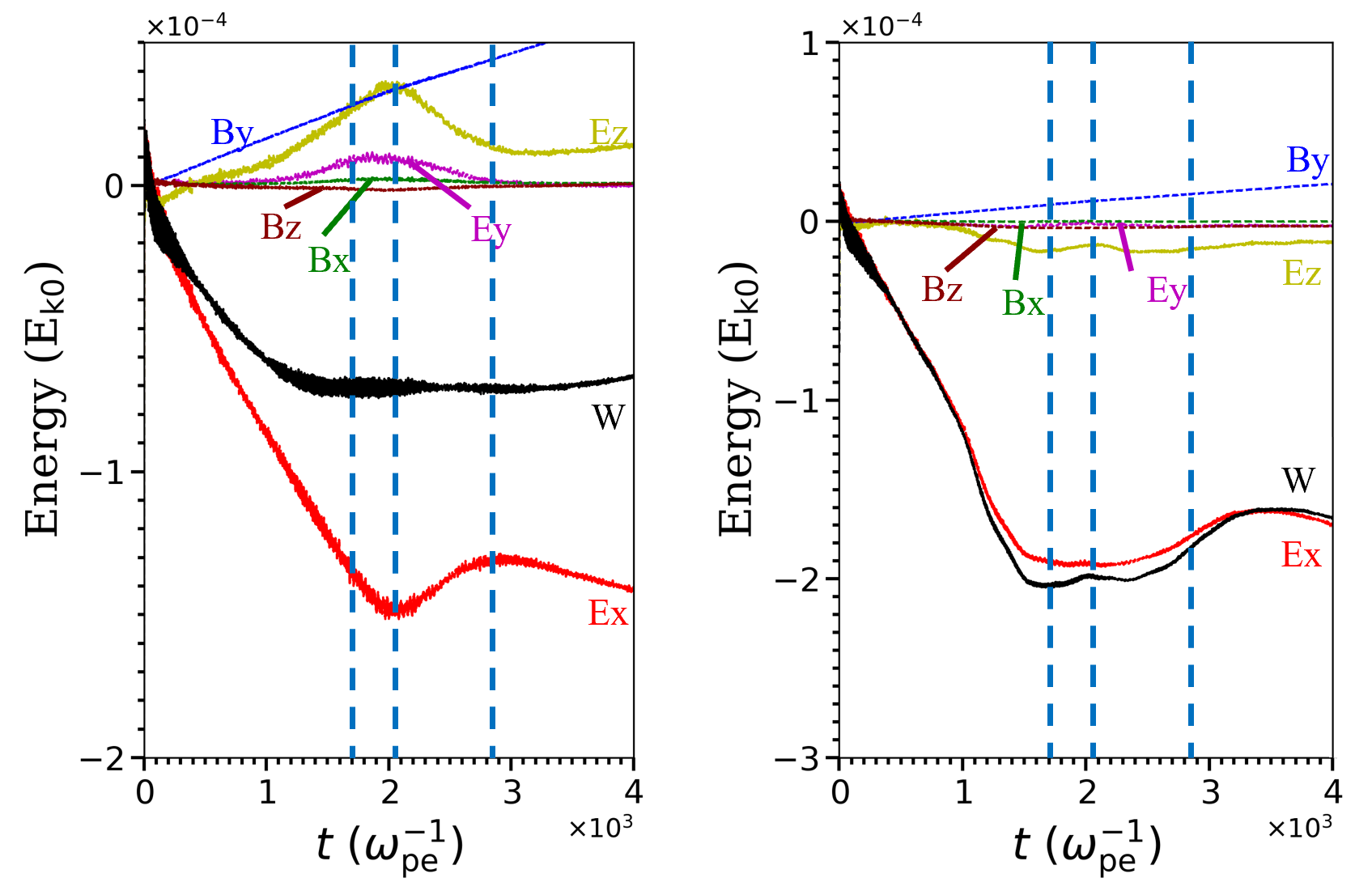}
\caption{Temporal evolution of EM field components ($E_x, E_y, E_z, B_x, B_y, B_z$) and total EM energy ($W$), normalized to their respective initial values. Left panel shows the full solution domain, and right panel shows the uniform region. Blue dashed lines demarcate four temporal phases: (1) Z mode propagation ($0-1700~\omega_\mathrm{pe}^{-1}$), (2) mode conversion ($1700-2100~\omega_\mathrm{pe}^{-1}$), (3) Z mode reflection ($2100-2900~\omega_\mathrm{pe}^{-1}$), (4) wave dissipation (after $2900~\omega_\mathrm{pe}^{-1}$).}
\label{fig:fields energy}
\end{figure}

According to Figure~\ref{fig:fields}, part of the downward-propagating Z mode gets reflected back into the domain due to the periodic boundary condition. This gets mixed with the newly-converted O mode signal and affects our estimates of the energy budget. Further studies can improve this by using a larger domain. Previous PIC simulations of nonlinear plasma emission processes \cite{Krafft_2021,Chen_2022,Zhang_2022}, show that only $\sim10^{-3}-10^{-4}$ of the energy of the beam-driven Langmuir turbulence can convert into escaping radiation. In our simulation, we started from directly pumping Z mode with appropriate conditions into the system, this explains why the mode conversion rate (~10-20\%) obtained here is much larger.

Saturn's NB emissions consist of two major components, one is at 20~kHz and the other is at 5~kHz. Both are attributed to the conversion of the Z mode waves upon density gradient. Yet, their source conditions exhibit significant differences in $\omega_\mathrm{pe}/\Omega_\mathrm{ce}$. For the 20~kHz one, $\omega_\mathrm{pe}/\Omega_\mathrm{ce}$ reaches $\sim5$, while for the 5~kHz one $\omega_\mathrm{pe}/\Omega_\mathrm{ce}$ is < 1 \cite{Ye_2010}. In addition, the conversion frequency for the 20~kHz one is $\sim\omega_{pe}$ and that for the 5~kHz one is $\sim\Omega_{ce}$. This may indicate that the two NB radiation originate from different conversion process. Further simulations are demanded to examine such difference. 

The mechanisms or the features or the observational characteristics of NB radiations vary across planets due to distinct source regions and plasma parameters. Table~\ref{tab:planets NB} collects published parameters for NB from Ganymede near its magnetopause ($\sim5.4~R_G$), from Earth near its plasmapause and from Jupiter at various sites: the extensive region (5-25~$R_J$) away from the plasma disk \cite{Reiner_1993,Boudouma_2024}, the day-side of the magnetosphere \cite{Gurnett_1983} and near the Jovian magnetopause \cite{Kaiser_1992}. Except for Saturn's 5~kHz NB radiation, $\omega_\mathrm{pe}/\Omega_\mathrm{ce}$ is always > 1 \cite{Kurth_1997,Gurnett_1983,Grimald_2008,Ye_2009,Wu_2021}. Further studies can explore the difference of conversion on different planets.

\section{Conclusions}\label{sec:conclusion}

We present the fully kinetic electromagnetic PIC simulations of the Z-to-O mode conversion induced by density gradient, using physical conditions in line with the Saturn's 20 kHz NB radiation. We started the simulation by pumping the Z mode wave with prescribed conditions into the domain, and simulated its further reflection and conversion into the quasi-parallel-propagating O mode emission. The simulation results agree with the predictions given by the linear magnetoionic wave theory of cold plasma and the Snell's law. We estimated the rate of energy conversion from Z mode to O mode to be ~10-20\%.

%
%

%
%

%
%
\newpage

\clearpage

\begin{table*}[hb]
	\centering
	\begin{tabular}{|p{2cm}|p{3cm}|p{3cm}|p{5cm}|}
		\hline
		~        & frequency (kHz) & $\omega_\mathrm{pe}/\Omega_\mathrm{ce}$ & Remarks on source location \\ \hline
		Saturn   & 5 or 20  & $\sim$5 for 20~kHz and $\sim$0.8 for 5~kHz   & Saturn’s plasma torus at L shells 8 to 10 for 5~kHz NB and L4 to 7 for 20~kHz NB \cite{Ye_2009} \\ \hline
		Jupiter  & 100-200  & $\sim$10    & mainly from the plasma disk within 5-25~$R_J$ \cite{Reiner_1993, Imai_2017, Boudouma_2024} \\ \hline
		Ganymede & 15-60    & $\sim$6     & Near Magnetopause, $\sim 5.4~R_G$ \cite{Kurth_1997} \\ \hline
		Earth    & 30-110   & $\sim$3-10  & Plasmapause boundary layer \cite{Grimald_2008} \\ \hline
	\end{tabular}
	\caption{Parameters of narrowband (NB) radiation from planetary magnetospheres.}
	\label{tab:planets NB}
\end{table*}
\clearpage

%
%
%
%

\appendix
\section{Linear theory and initial conditions}\label{sec:LMCT}

Equation~\eqref{eq1} presents the dispersion relation of cold plasma \cite{Kalaee_2020}. where $n_\parallel$ and $n_\perp$ denote the components of refractive index parallel and perpendicular to $\vec{B}_0$. $\varGamma = \left[ Y^4 \sin^4 \theta + 4(X-1)^2 Y^2 \cos^2 \theta \right]^{\frac{1}{2}}, X = \left( \dfrac{\omega_p}{\omega} \right)^2, Y = \dfrac{\omega_c}{\omega}$, where $\omega_c$ is the electron frequency, $\theta$ is the wave normal angle, and the +(-) sign indicates the refractive index of the extraordinary (ordinary) mode. The ordinary and extraordinary modes coalesce when $\varGamma=0$, requiring $X=1$ and $\theta=0$. Under these conditions, Equation~\eqref{eq1} simplifies to $n^2 = n_\parallel^2 = \frac{Y}{1 + Y}$, indicating that mode conversion occurs at $\omega_p\sim\omega$ with $\theta=0$ and $n_\perp=0$. We calculated the plasma frequency at the reflection point ($\omega_r$) for a given wave frequency ($\omega$) using Equations 18 and 19 from \citeA{Kalaee_2020}.

\begin{align}
	n^2 &= n_\parallel^2 + n_\perp^2 = 1 - \frac{2X(X-1)}{2(X-1) - Y^2 \sin^2 \theta \pm \Gamma}. \label{eq1}
\end{align}

Applying the Snell’s law to Saturn’s 20 kHz NB emission, we can derive the parameters of the to-be-pumped Z mode in the uniform region ($\omega = 5.03~\Omega_\mathrm{ce}$) that are angle $\theta_\mathrm{0}=70.6^{\circ}$ and $k_\mathrm{0} = 6.0~\Omega_\mathrm{ce}/c$. By solving the dispersion relation and the Snell's law, we trace the Z mode propagation in the density gradient region (see Figure~\ref{fig:Z mode solution}). As the Z mode propagates into denser region, $\theta$ decreases from $70^{\circ}$ to $\sim 30^{\circ}$, then it reflects back into the dilute region with $\theta$ changing from $30^{\circ}$ to $-30^{\circ}$. Then $\theta$ approaches $0^{\circ}$ causing the conversion of Z mode into O mode.

\section{Wave-pumping setup}\label{sec:pumping}

Based on the linear cold plasma wave theory, we can determine the relative strengths of the six components of the electromagnetic field for any given mode with known parameters of $\omega$, $k$, and $\theta$ \cite{Ni_2021}. This prescribes the field components of the Z mode to-be-pumped, according to the following equations, where phase $\Phi = k \cdot \sin \theta \cdot x + k \cdot \cos \theta \cdot z $.

\begin{align}
	\vec{E} &= E_x \sin(\Phi) \hat{e}_x + E_y \cos(\Phi) \hat{e}_y + E_z \sin(\Phi) \hat{e}_z, \\
	\vec{B} &= B_x \cos(\Phi) \hat{e}_x + B_y \sin(\Phi) \hat{e}_y + B_z \cos(\Phi) \hat{e}_z, \\
	E_y &= 0.0001, \\
	E_x &= \frac{(N^2 - S)}{D} \cdot E_y, \\
	E_z &= \frac{(S-N^2) \cdot (S-N^2 \cos^2\theta)}{ DN^2 \sin \theta \cos \theta } \cdot E_y, \\
	B_x &= - \frac{k \cos \theta}{\omega} \cdot E_y, \\
	B_y &= \left(\frac{k \cos \theta}{\omega} \cdot E_x - \frac{k \sin \theta}{\omega} \cdot E_z\right) \cdot E_y, \\
	B_z &= \frac{k \sin \theta}{\omega} \cdot E_y, \\
	S   &= 1 - \frac{\omega_\mathrm{pe}^{2}}{\omega^2-\Omega_\mathrm{ce}^2}, \\
	N   &= \frac{kc}{\omega}, \\
	D   &= - \frac{\Omega_\mathrm{ce} \cdot \omega_\mathrm{pe}^2}{\omega \cdot \left(\omega^2 - \Omega_\mathrm{ce}^2 \right)}.
\end{align}

\section{The method of mode identification and estimate of conversion efficiency}\label{sec:con eff}

One can identify wave modes with dispersion diagrams given by the Fourier analysis \cite{Chen_2022,Zhang_2023}. Since the Z and O modes are close in dispersion diagrams at the condition of mode conversion, it is challenging to distinguish them according to the dispersion diagrams. Here, we do this by calculating the ratios of various electric field components according to the above expressions. The results are presented in Figure~\ref{fig:theory mode}. The amplitudes of $E_x$ and $E_z$ of the Z mode decrease with decreasing $\theta$. For quasi-parallel propagation, $E_z$ of the O mode is stronger than that of the Z mode. Therefore, a significant enhancement of $E_z$ indicates the presence of the O mode.

\begin{figure}
	\noindent
	\includegraphics[width=0.8\textwidth]{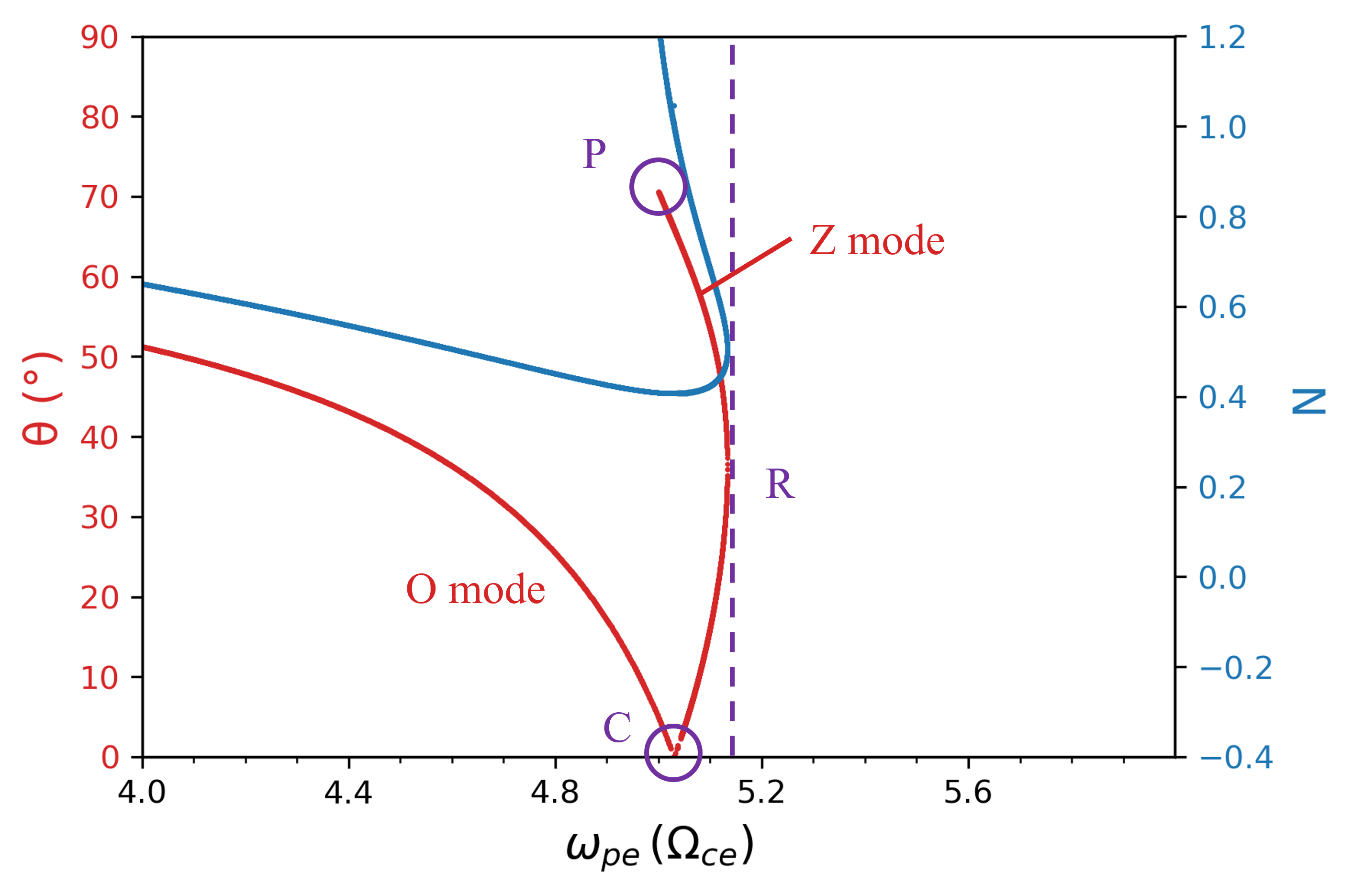}
	\caption{Ray-tracing of Z mode propagation in density gradient. X-axis: local electron plasma frequency normalized by electron gyro-frequency ($\Omega_\mathrm{ce}$). Curves show evolution of propagation angle $\theta$ (red line) and refractive index $N$ (blue line). Labels: P (pumping), R (return), C (conversion).}
	\label{fig:Z mode solution}
\end{figure}

\begin{figure}
	\noindent
	\includegraphics[width=0.8\textwidth]{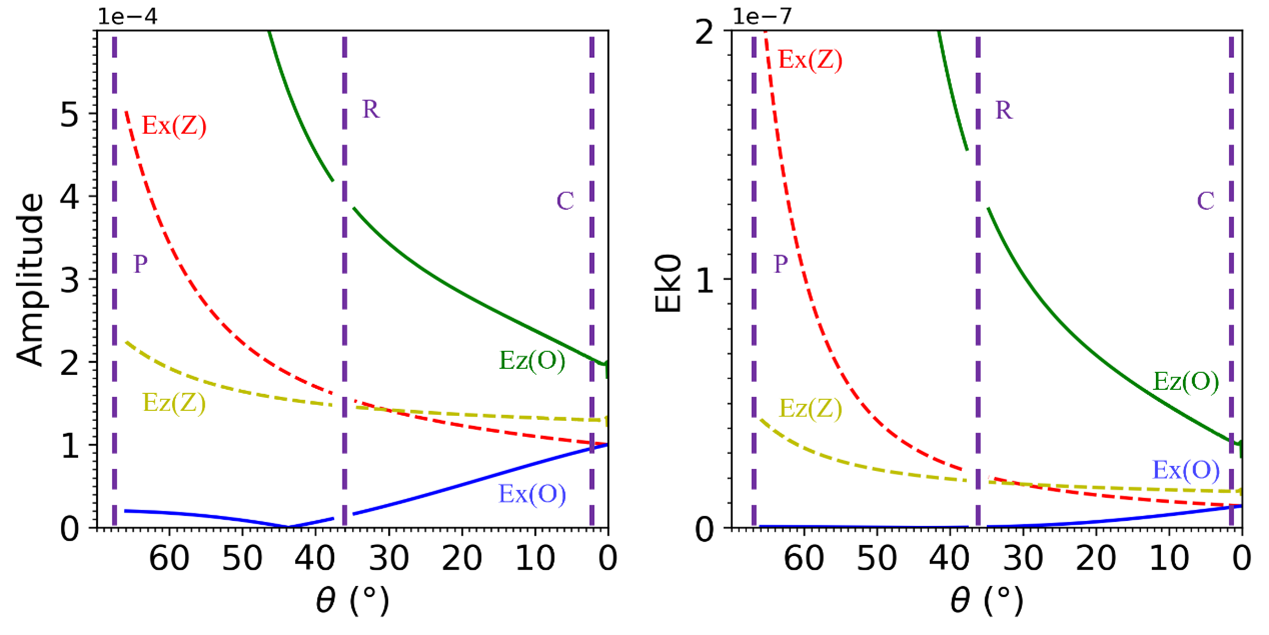}
	\caption{Amplitude (left) and energy (right) profiles of $E_x$ and $E_z$ components for Z mode (dashed) and O mode (solid) along the propagation path, calculated using cold-plasma theory. Data gaps correspond to the return zone. Labels: P (pumping), R (return), C (conversion).}
	\label{fig:theory mode}
\end{figure}

According to the linear theory, the O mode leaves the conversion regime at a quasi-parallel propagation, whereas the leftover Z mode leaves with a large propagation angle (see Figure~\ref{fig:Z mode solution}). As shown by the dispersion curves (Figure~\ref{fig:theory mode}), the O mode propagates much faster in group velocity that that of the Z mode. This means the dominance of the O mode in the uniform region shortly after the conversion. We calculated the electric field energies for Z mode ($W_\mathrm{E}^\mathrm{Z}$) and O mode ($W_\mathrm{E}^\mathrm{O}$) within the black-dashed region of Figure~\ref{fig:fields} to determine the rate of mode conversion ($\eta = \left( W_\mathrm{E}^\mathrm{O} \right) / \left( W_\mathrm{E}^\mathrm{Z} \right)$). As shown in Figure~\ref{fig:fields}, a portion of the downward-propagating Z mode was reflected back into the domain, resulting in an enhancement of the $E_x$ component in the uniform region. We excluded this enhanced component when calculating the mode energy.

\newpage

\begin{figure}
	\noindent
	\includegraphics[width=0.8\textwidth]{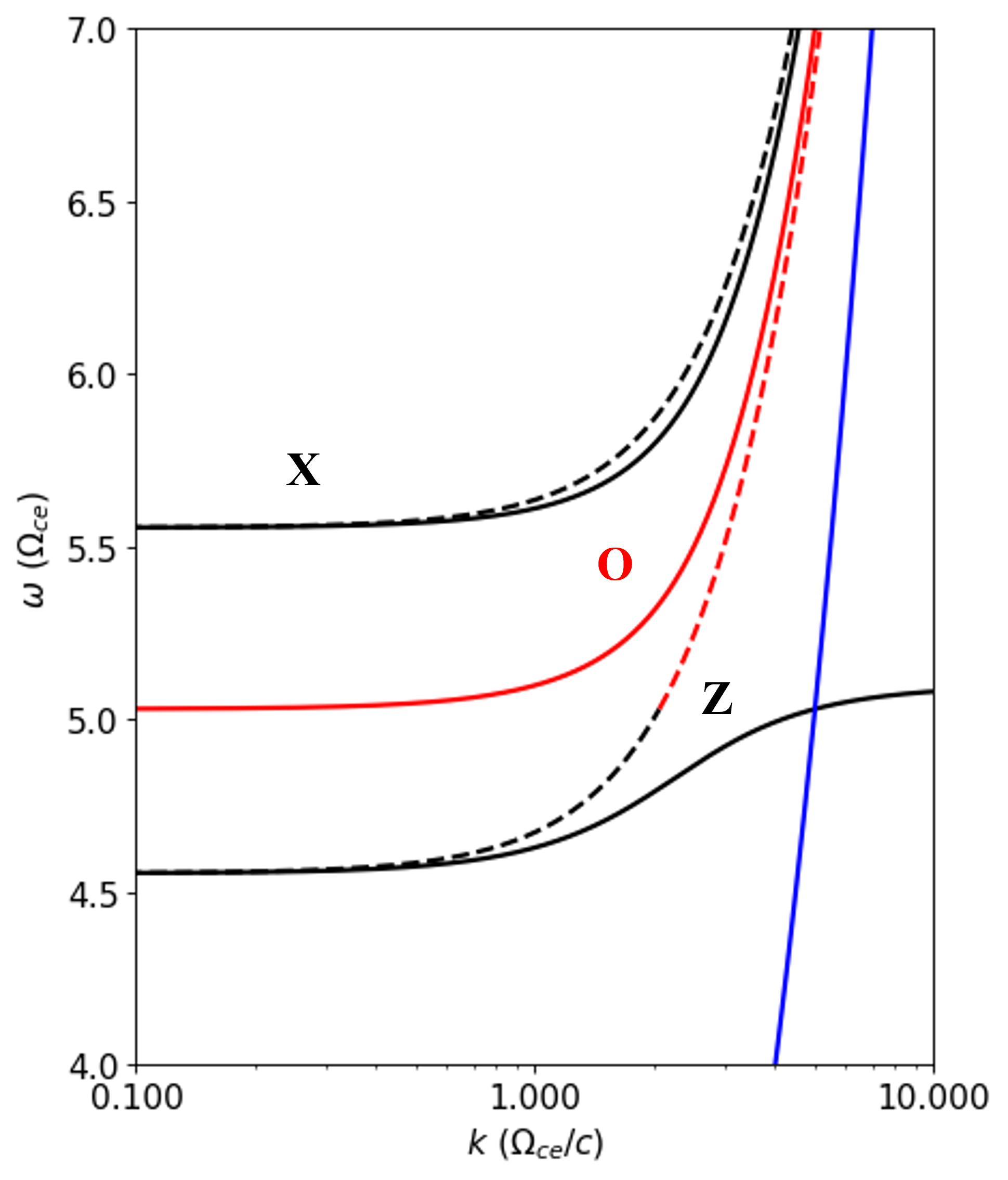}
	\caption{Theoretical dispersion relation at $\omega = 5.03~\Omega_\mathrm{ce}$. The red black and blue lines mark O mode Z mode and light wave ($\omega = kc$).The solid and dashed lines represent $\theta = 70^\circ$ and $\theta = 0^\circ$, The slope of the curve gives the group velocity.}
	\label{fig:theory dispersion relation}
\end{figure}
\clearpage

\section*{Open Research Section}
All data necessary to validate the findings presented in this manuscript can be found by \citeA{Mu_2024}.

\acknowledgments
This study is supported by the National Natural Science Foundation of China (NNSFC) grants (Nos. 12103029, 12303061, 12203031), Shandong Provincial Natural Science Foundation (ZR2023QA141). The authors acknowledge the Beijing Super Cloud Computing Center (BSC-C, URL: http://www.blsc.cn/) for providing HPC resources, and the open-source Vector-PIC (VPIC) code provided by Los Alamos National Labs (LANL). We thank Professor Ye Shengyi for providing the observational data of Saturn's narrowband radiation, and Professors Lu Quanming and Li Gang for their constructive suggestions. We also appreciate Professor Yuto Katoh's assistance with the electron fluid model.

%
%
\bibliography{CAFSaturn20kHz} 

%
%
%
%
%

\end{document}